\documentclass[prd,twocolumn,superscriptaddress,amsmath]{revtex4-1}
\usepackage{graphicx}
\usepackage{bm}
\usepackage{slashed}
\usepackage{color}
\usepackage[colorlinks,citecolor=blue,urlcolor=blue,linkcolor=blue]{hyperref}

\begin{document}
	
	\title{Unveiling CP property of top-Higgs coupling with graph neural networks at the LHC}
	
	\author{Jie Ren}
	\email{renjie@itp.ac.cn}	
	\affiliation{CAS Key Laboratory of Theoretical Physics, Institute of Theoretical Physics, Chinese Academy of Sciences, Beijing 100190, China}
	\affiliation{School of Physics, University of Chinese Academy of Sciences, Beijing 100049, China}
	
	\author{Lei Wu}
	\email{leiwu@itp.ac.cn}	
	\affiliation{School of Physics Science and Technology, Nanjing Normal University, Nanjing, 210023, China}
	
	\author{Jin Min Yang}
	\email{jmyang@itp.ac.cn}
	\affiliation{CAS Key Laboratory of Theoretical Physics, Institute of Theoretical Physics, Chinese Academy of Sciences, Beijing 100190, China}
	\affiliation{School of Physics, University of Chinese Academy of Sciences, Beijing 100049, China}
	\affiliation{Department of Physics, Tohoku University, Sendai 980-8578, Japan}
	
	
	\begin{abstract}
		The top-Higgs coupling plays an important role in particle physics and cosmology. The precision measurements of this coupling can provide an insight to new physics beyond the Standard Model. In this paper, we propose to use Message Passing Neural Network (MPNN) to reveal the CP nature of top-Higgs interaction through semi-leptonic channel $pp \to t(\to b\ell^-\nu_\ell)\bar{t}(\to \bar{b}jj)h(\to b\bar{b})$. Using the test statistics constructed from the event classification probabilities given by the MPNN, we find that the pure CP-even and CP-odd components can be well distinguished at the LHC, with at most 300~fb$^{-1}$ experimental data.
	\end{abstract}
	
	\maketitle
	
	
	\section{Introduction}
	
	The discovery of Higgs boson~\cite{Aad:2012tfa, Chatrchyan:2012xdj} is a great step in the long quest to the origin of mass. The precision measurement of Higgs couplings is one of the main goals of the future LHC experiment, which will further reveal the electroweak symmetry breaking mechanism and shed light on new physics beyond the Standard Model (SM). Among these couplings, the top-Higgs interaction is particularly interesting. On one hand, due to its large size, this coupling dominantly contributes to the renormalization group evolution of the Higgs potential and thus plays an unique role in determining the scale of new physics. On the other hand, the direct extraction of this coupling from the QCD process $pp \to t\bar{t}H$ is very challenging since it has a small production rate and rather complicated final states. Very recently, the ATLAS and CMS collaborations have reported the observation of $t\bar{t}H$ production in several Higgs decay channels, such as $\gamma\gamma$ and $W^+W^-$, at the 13~TeV LHC~\cite{Aaboud:2018urx, Sirunyan:2018hoz}. The LHC Run-3 with a higher luminosity will have great potential to decipher the structure of top-Higgs coupling.

	The general top-Higgs interaction can be parameterized as
	\begin{equation}
		{\cal L} \supset -\frac{y_t}{\sqrt{2}} \bar{t}(\cos\xi + i \gamma_5 \sin\xi) t H,
		\label{tth-coupling}
	\end{equation}
	where $y_t = \sqrt{2} m_t / v$ and $\xi = 0$ in the SM~\cite{AguilarSaavedra:2009mx}, with $v = 174$~GeV being the vacuum expectation value of the Higgs field. The presence of $\sin\xi$ term leads to the CP violation in top-Higgs coupling. It will affect the Higgs production and decay channels, the electric dipole moments and the flavor physics observables, which are measured to well agree with the SM and then set strong constraints on the CP-violating top-Higgs coupling~\cite{Cirigliano:2016njn,Kobakhidze:2016mfx}. However, these indirect bounds are model-dependent and can be evaded in some extensions of the SM. Therefore, the most robust test of top-Higgs coupling in Eq.~(\ref{tth-coupling}) is from the direct measurement of $t\bar{t}H$ production at colliders. Several observables, such as $t\bar{t}$ spin correlation and charge asymmetry in $t\bar{t}H$ production, are constructed to probe the CP violating top-Higgs coupling at the LHC~\cite{Gunion:1996xu, Ellis:2013yxa, Bramante:2014gda, Demartin:2014fia, Aguilar-Saavedra:2014kpa, Godbole:2015bda, Buckley:2015vsa, Li:2015kaa, Li:2017dyz, Cao:2018}.

	On the other hand, applying machine learning (ML) techniques, exceptional performance can be achieved with object level kinematic variables from jets, leptons, and photons to separate the signal from the background~\cite{Bertone:2016mdy, Baldi:2014kfa, Baldi:2014pta, Bridges:2010de, Buckley:2011kc, Bornhauser:2013aya, Caron:2016hib}. A successful example in this aspect is the use of boosted decision trees in the LHC experiment that led to the discovery of Higgs boson~\cite{Roe:2004na}. Since a collision event can be seen as a geometrical pattern formed by a number of final state objects, graph is a natural way to represent the events in mathematical language, which can be efficiently analyzed by an appropriate ML approach. Among the ML algorithms to deal with graphs, the Message Passing Neural Networks (MPNNs)~\cite{2017arXiv170401212G} are particularly suited for graph classification and flexible enough as the original Graph Neural Networks (GNNs)~\cite{Gori05, Scarselli09} as nonlinear end-to-end models that relate the target output to the input graphs. An MPNN consists of a number of learnable functions acting on the graph nodes, and can be efficiently trained using supervised learning techniques. So far, the MPNNs have been successfully applied in supersymmetry exploration~\cite{Abdughani:2018wrw}, jet physics~\cite{Henrion17} and other fields~\cite{2017arXiv170401212G}.

	In this paper, we attempt to use MPNN to investigate the CP nature of top-Higgs coupling in Eq.~(\ref{tth-coupling}). We design and train a specific MPNN to classify the collider events, and then perform a hypothesis test based on the variable constructed from the output of the MPNN. As a proof of concept, we focus on the semi-leptonic top decay channel of the process $pp \to t\bar{t} H(\to b\bar{b})$ at the LHC.	
	

	\section{Methods}
	
	For convenience, we denote the Higgs boson with CP-even $t\bar{t}H$ coupling ($\xi = 0$) as $h$, and the one with CP-odd $t\bar{t}H$ coupling ($\xi = \pi/2$) as $A$. At the LHC, the $t\bar{t}h$ and $t\bar{t}A$ signals have the same background events dominantly coming from the process $pp \to t\bar{t}b\bar{b}$.
	
	We choose event graphs~\cite{Abdughani:2018wrw} as the representation of collider events, and design an MPNN specific to the classification of the collider events, whose outputs are the probabilities of the input event graph being $t\bar{t}h$, $t\bar{t}A$ and $t\bar{t}b\bar{b}$ event, respectively. Then, we construct a variable from the output of MPNN and perform a hypothetical test.
	
	\subsection{Event graph}
	
	\begin{figure}[t]
		\includegraphics[width=8cm,height=8cm]{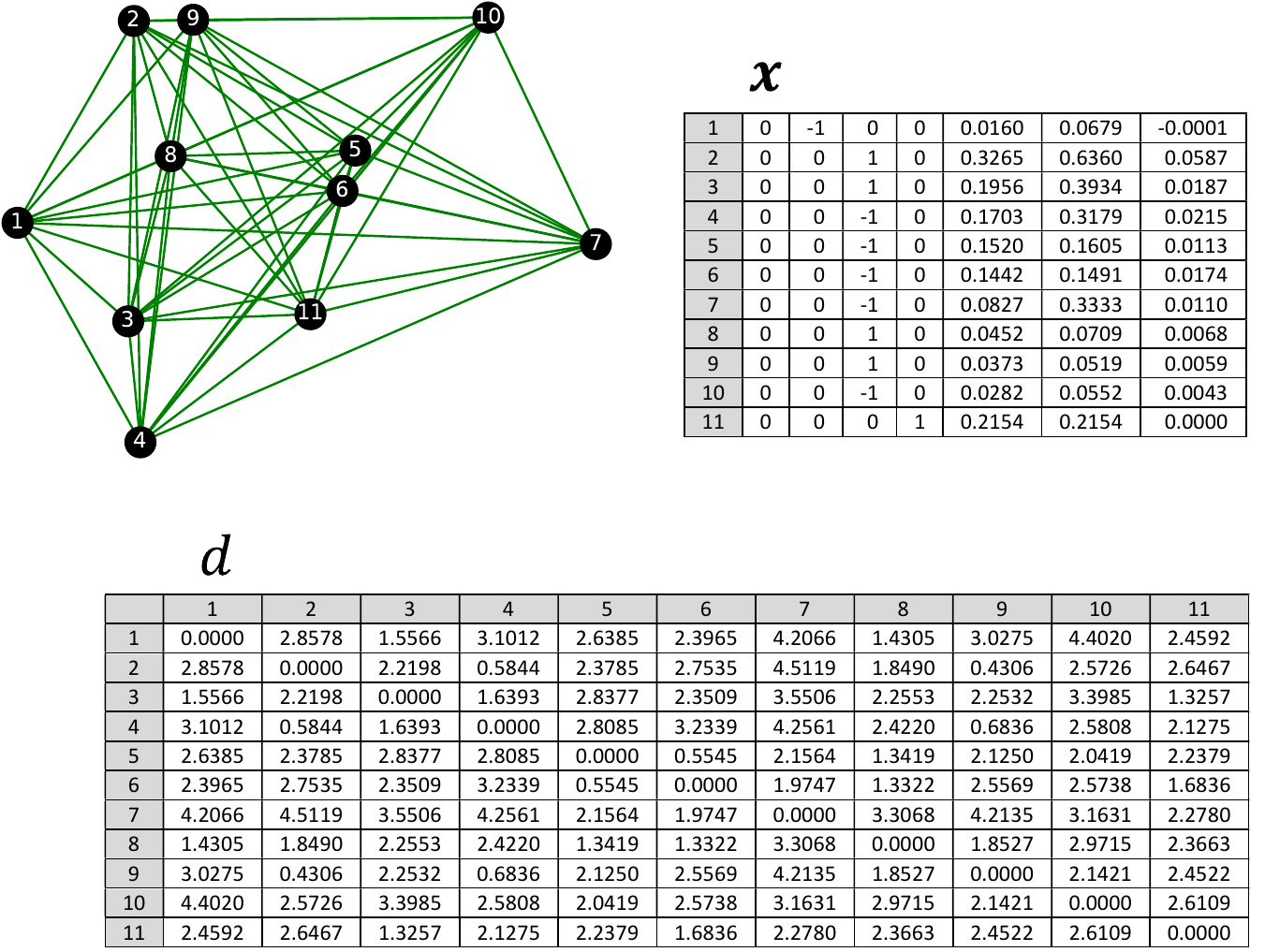}
		\caption{An event graph with detailed node features and edge weights for a specific simulated $t\bar{t}h$ event.}
		\label{event-graph}
	\end{figure}
	
	As the input of MPNN, we represent each of the collider events as an event graph. FIG.~\ref{event-graph}, as illustration, shows an event graph for a specific simulated $t\bar{t}h$ event.
	
	For a given collider event, the nodes in the graph are used to represent the final state objects, including the reconstructed photons, leptons, jets and missing transverse momentum (MET). Each node has a compact seven dimensional feature vector $\bm{x}_i = (I_1, I_2, I_3, I_4, p_T, E, m)$ to describe the major properties of the corresponding final state. Except the transverse momentum $p_T$, energy $E$ and mass $m$, the first four features are indicators of the type of final state: (1) it is a photon ($I_1 = 1$) or not ($I_1 = 0$); (2) it is a lepton ($I_2 = $charge) or not ($I_2 = 0$); (3) It is a $b$-jet ($I_3 = 1$), light jet ($I_3 = -1$) or not a jet ($I_3 = 0$); (4) it is the MET ($I_4 = 1$) or not ($I_4 = 0$).
	
	Each pair of nodes is linked by an edge, which is weighted by the geometrical distance between the corresponding pair of final states. We choose to use $d_{ij} = \sqrt{\Delta(\eta_i, \eta_j)^2 + \Delta(\phi_i, \phi_j)^2}$ to measure the geometrical distance between two final states $i$ and $j$, where $\eta$ and $\phi$ are the pseudo-rapidity and azimuthal angle, respectively.
	
	Notice that the differential cross section of a collider event is invariant with the rotation of the whole event along the beam. To respect such an important geometrical symmetry of collider event, we exclude the information of azimuthal angle from the node features, and only encode the difference of azimuthal angles in the edge weights. In such a design, the event representation and classification will be stable, regardless of the rotation of event along the beams. Note also that, (1) the number of nodes in an event graph depends on specific collider event, (2) there is no ordering of nodes, and (3) the data in event graphs are exact. These are the main differences between event graph representation and other collider event representations used as input for ML models.	
	
	\begin{figure}[t]
		\includegraphics[width=8cm,height=5cm]{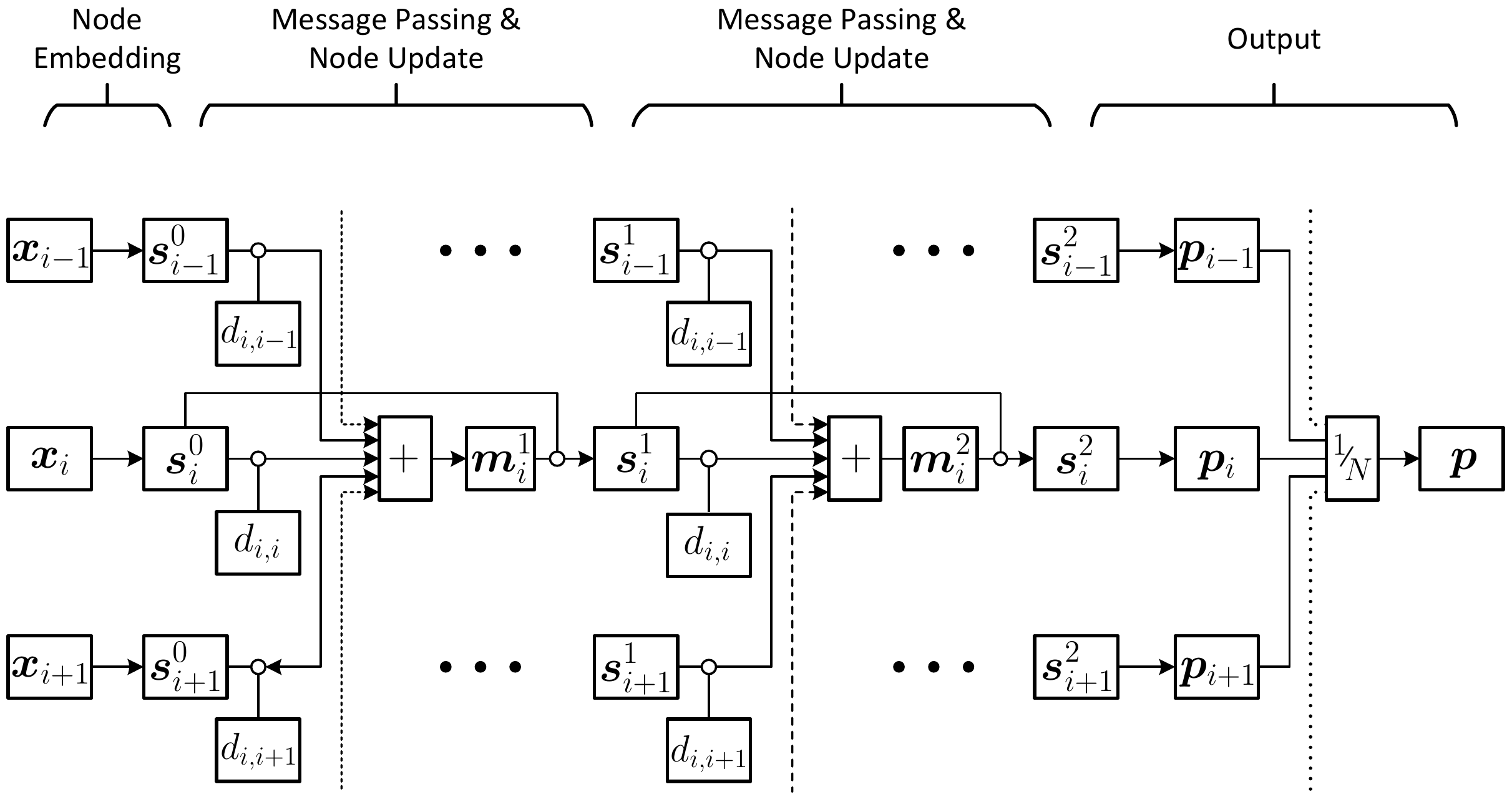}
		\caption{The architecture of MPNN designed for classifying $t\bar{t}h$, $t\bar{t}A$ and $t\bar{t}b\bar{b}$ events. It has one node embedding layer, two message passing and node update layers and one output layer. The small circles denote vector concatenation. The arrows denote applying non-linear functions. The summation and average run over all nodes.}
		\label{mpnn-arch}
	\end{figure}
	
	\subsection{Network architecture}
	
	The architecture of our MPNN is shown in FIG.~\ref{mpnn-arch}. It has one node embedding layer, two message passing and node update layers and one output layer.
	
	The node embedding layer embeds each node feature vector $\bm{x}_i$ into a higher dimensional node state vector $\bm{s}_i^0$ by applying a linear transformation and the rectified linear unit (ReLU) activation function,
	\begin{equation}
		\bm{s}_i^0 = \mathrm{ReLU}(W_e \bm{x}_i + \bm{b}_e) ,
	\end{equation}
	where $W_e$ and $\bm{b}_e$ are learnable parameters. The state vector $\bm{s}_i^0$ only encodes the node features $\bm{x}_i$ without any information about the geometrical pattern of the graph.
	
	In the following two message passing and node update layers, the nodes exchange information contained in their state vectors by passing messages. At layer $t$, each node $i$ collects the messages sent from each nodes $j$ and then update its state vector,
	\begin{eqnarray}
		\bm{m}_i^t &=& \sum_j \bm{m}_{i \leftarrow j}^t = \sum_j \mathrm{ReLU}(W_m^t [\bm{s}_j; \hat{\bm{d}}_{ij}] + \bm{b}_m^t) , \label{message-func} \\
		\bm{s}_i^t &=& \mathrm{ReLU}(W_u^t [\bm{s}_i, \bm{m}_i] + \bm{b}_u^t) ,
	\end{eqnarray}
	where the brackets denote vector concatenation, $W$s and $\bm{b}$s are learnable parameters in each layer. Note that, to make the edge weight $d_{ij}$ more suitable in linear transformation, we expand it onto 21 Gaussian bases to form a weight vector $\hat{\bm{d}}_{ij}$, whose components are
	\begin{equation}
		(\hat{\bm{d}}_{ij})_k = \exp \left\{ \frac{(d_{ij} - \mu_k)^2}{2 \sigma^2} \right\} ,
	\end{equation}
	where $\mu_k$ are linearly distributed in range [0, 5] and $\sigma = 0.25$. The message passing mechanism is the key to automatically extract features of the input event graph, which can efficiently disseminate the information among all the nodes taking into account the connections between the nodes. After the two message passing and node update layers, each node state vector can be viewed as an encoding of the whole event graph.
	
	In the output layer, each node $i$ produces three probabilities $\bm{p}_i$ by applying a linear transformation and the softmax activation function on its state vector $\bm{s}_i$,
	\begin{equation}
		(\bm{p}_i)_k = \frac{\exp \{ (W_o \bm{s}_i + \bm{b}_o)_k \} }{\sum_k \exp \{ (W_o \bm{s}_i + \bm{b}_o)_k \} }, \quad (k = 1,2,3)
	\end{equation}
	To stablize the classification performance, we average the output over all the nodes as the final output of MPNN,
	\begin{equation}
		\bm{p} = \frac{1}{N} \sum_i \bm{p}_i ,
	\end{equation}
	where $N$ is the number of nodes in the input event graph. The three components of $p$ are the probabilities of the single input event graph $e$ being the $t\bar{t}h$, $t\bar{t}A$ and $t\bar{t}b\bar{b}$ event, respectively, denoted as $p(h|e)$, $p(A|e)$ and $p(b|e)$.
	
	It is worth to note that the MPNN is a dynamic neural network, which can be viewed as a stack of several learnable transformations acting on each signal or pair of graph nodes. Therefore, MPNN intrinsically scales with the size of input event graph.
	
	\subsection{Training}
	
	The MPNN can be efficiently trained using supervised learning. We choose cross entropy as the loss function. The gradients of loss to learnable parameters are evaluated on each mini-batch of 500 examples. The learnable parameters are optimized using the ADAM optimizer~\cite{KingmaB14} with a fixed learning rate of 0.001. The training is performed up to 300 epochs and we choose the MPNN parameters which lead to the best generalization performance (minimum loss) on the validation set. All the above are implemented with the open-source deep learning framework PyTorch~\cite{pytorch} with extensive GPU acceleration.
	
	\subsection{Hypothesis test}

	If the top-Higgs coupling is CP-even, the event sample collected in experiments will come from the $t\bar{t}h$ process plus the dominate $t\bar{t}b\bar{b}$ background process. Otherwise, it consists of a mixing of $t\bar{t}A$ and $t\bar{t}b\bar{b}$ events. Therefore, we define variables that can discriminate event samples of the two scenarios. From the single-event probabilities output from the MPNN, we construct two likelihoods
	\begin{eqnarray}
		L_h(D) &=& {\prod_{e \in D}}' p(h|e) \\
		L_A(D) &=& {\prod_{e \in D}}' p(A|e)
	\end{eqnarray}
	to measure the consistence of a given event sample $D$ with each of the two scenarios. In the CP-even scenario, $L_h(D) \gg L_A(D)$; otherwise in the CP-odd scenario, $L_h(D) \ll L_A(D)$. It is worth to note that the productions only run over events with $p(h|e)$ and $p(A|e)$ larger than $p(b|e)$. Namely, we exclude the background-like events in the evaluation, which can effectively reduce the contamination of background.

	To perform a hypothesis test, here we choose to use the log-likelihood ratio
	\begin{equation}
		\ln Q(D) = \ln \frac{L_A(D)}{L_h(D)}
	\end{equation}
	as the test statistics. The distribution of $\ln Q$ in the two scenarios, denoted as $f_h$ and $f_A$, respectively, can be numerically obtained by evaluating a large number of random simulated event samples, namely, performing pseudo experiments.
	
	\begin{figure}[t]
		\includegraphics[width=6cm]{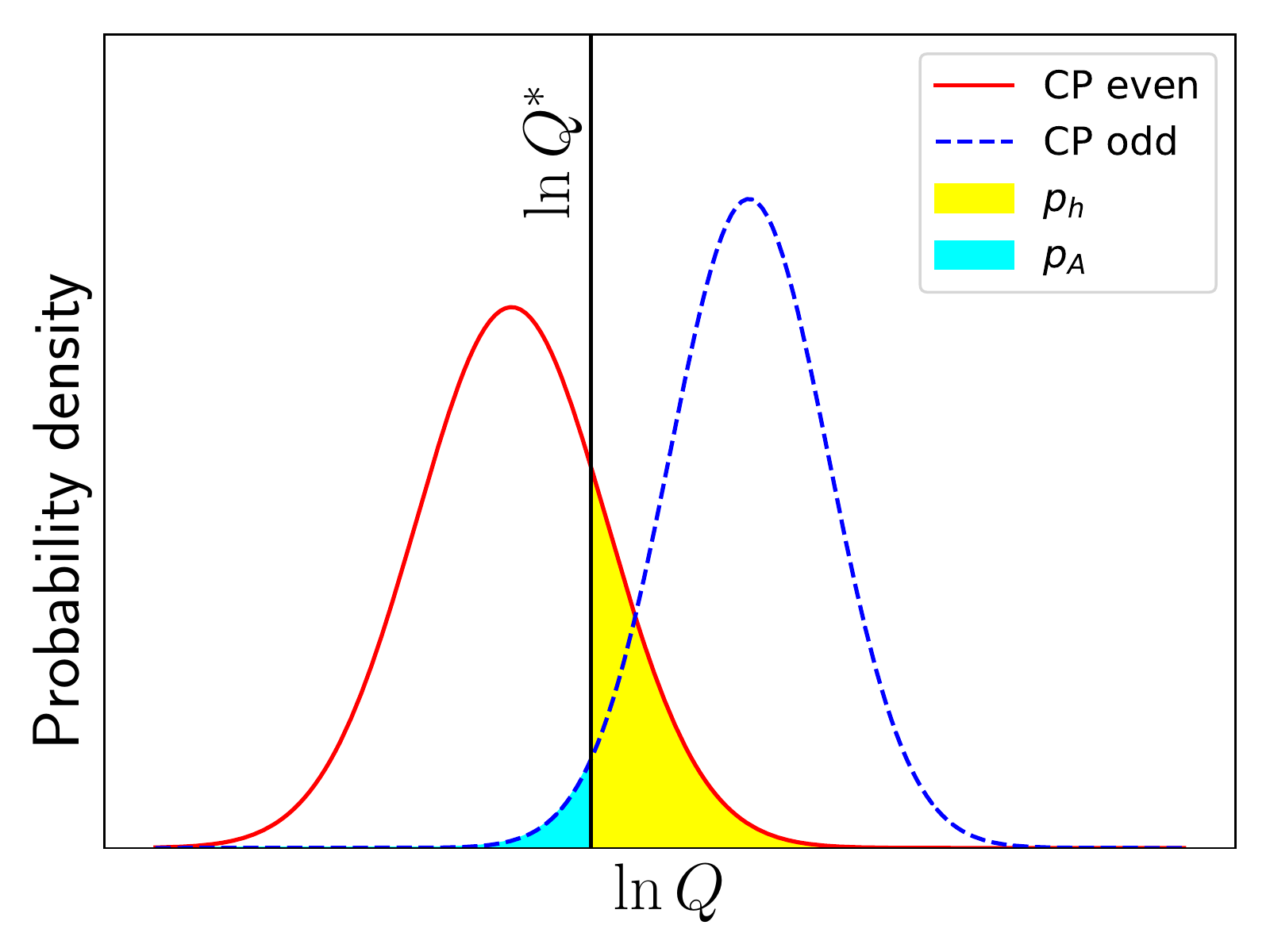}
		\caption{Illustration of the evaluation of $p$-values of rejecting the CP-even scenario and the CP-odd scenario, respectively.}
		\label{p-value-calc}
	\end{figure}	
	
	Because actually generating a huge number of simulated events can be extremely time-consuming, we adopt the bootstrap technique. First, for each process $X$, we generate a simulated event dataset, in which the number of events is large enough and the events have equal weights. Then, we construct event samples of process $X$ by randomly sampling the corresponding dataset with replacement. The number of events $n$ in an event sample obeys Poisson distribution $P(n | \lambda)$ with the average number of events $\lambda = \epsilon_X \sigma_X L$, where $\sigma_X$ is the production cross section of process $X$, $L$ is the integrated luminosity and $\epsilon_X$ is the event selection efficiency for process $X$. In the CP-even (odd) scenario, a pseudo experimental event samples is the union of a $t\bar{t}h$ ($t\bar{t}A$) sample and a $t\bar{t}b\bar{b}$ sample.
	
	Given the distributions of $\ln Q$ in the two scenarios and the $\ln Q^*$ calculated from the observed experimental data $D^*$, as shown in FIG.~\ref{p-value-calc}, we can evaluate the $p$-values of rejecting the CP-even scenario and the CP-odd scenario by the integrals
	\begin{eqnarray}
		p_h(\ln Q^*) &=& \int_{\ln Q^*}^{+\infty} f_h(x) dx \\
		p_A(\ln Q^*) &=& \int_{-\infty}^{\ln Q^*} f_A(x) dx
	\end{eqnarray}
	respectively.
	
	\section{Results}
	\begin{figure}[t]
		\includegraphics[width=8cm]{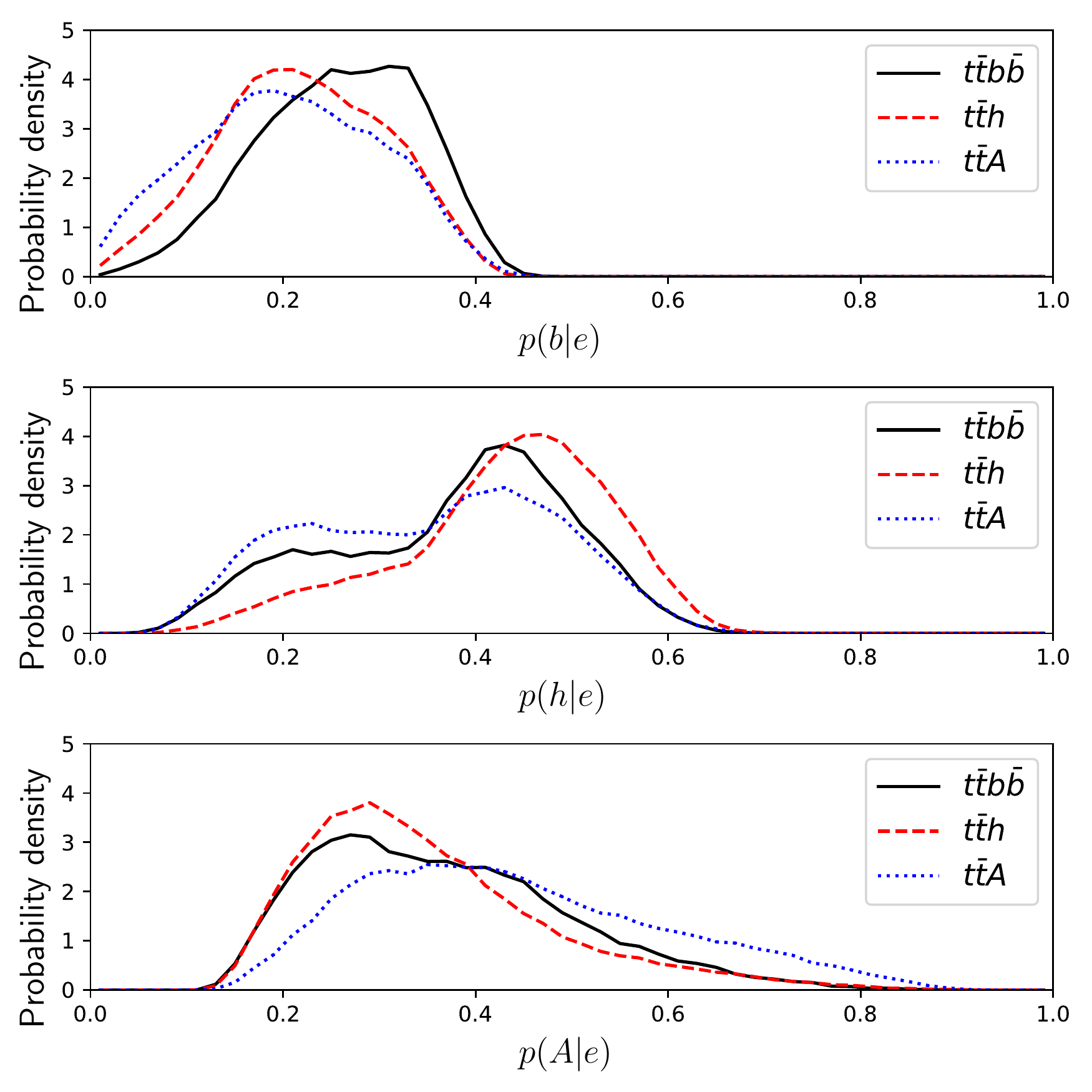}
		\caption{The distribution of MPNN output $p(h|e)$, $p(A|e)$ and $p(b|e)$ for $t\bar{t}h$, $t\bar{t}A$ and $t\bar{t}b\bar{b}$ events, respectively.}
		\label{p-dist}
	\end{figure}
	Simulated events of the $t\bar{t}h$, $t\bar{t}A$ and $t\bar{t}b\bar{b}$ processes are generated separately using MadGraph5~\cite{Alwall:2014hca} at 13~TeV LHC. Showering and hadronization are performed by Pythia8~\cite{Sjostrand:2014zea}. Detector simulation is done by Delphes~\cite{deFavereau:2013fsa} with ATLAS configuration. CheckMATE2~\cite{Drees:2013wra} is used to perform event selection. Leptons are detected within $p_T > 20$~GeV and $|\eta| < 2.5$, and jets are required to have $p_T > 25$~GeV and $|\eta| < 2.5$. B-tagging is performed with 60\% nominal efficiency. We focus on the semi-leptonic channel, requiring events to have exactly one lepton, four $b$-jets and at least two light jets in the final states.
	
	After event selection, the detection cross sections $\epsilon\sigma$ for the $t\bar{t}h$, $t\bar{t}A$ and $t\bar{t}b\bar{b}$ processes are 3.78, 1.82 and 27.5~fb, respectively. We collect 900,000 examples with balanced number of $t\bar{t}h$, $t\bar{t}A$ and $t\bar{t}b\bar{b}$ events as the training set for optimizing the parameters in our MPNN, while another 300,000 examples are collected as the validation set for performance evaluation.

	We show in FIG.~\ref{p-dist} the distributions of the output of our trained MPNN evaluated on the validation set. It is clear that the MPNN has successfully learned important discriminative event features for different processes. The background events are prone to have higher $p(b|e)$, while the $t\bar{t}h$, $t\bar{t}A$ events get higher $p(h|e)$ and $p(A|e)$, respectively.
	
	\begin{figure}[t]
		\includegraphics[width=8cm]{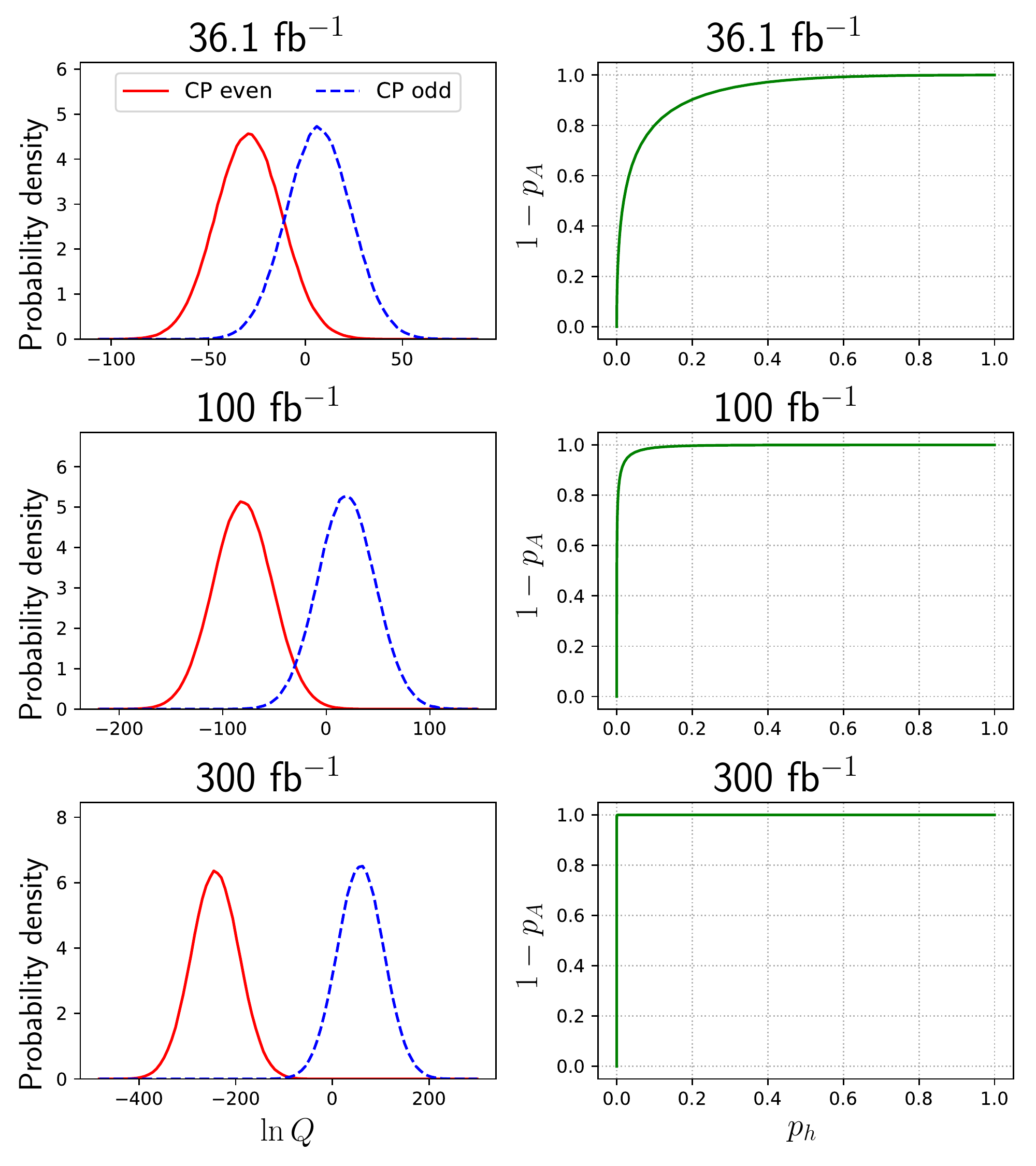}
		\caption{(left) The distribution of log-likelihood ratio from pseudo experiments in the CP-even scenario and the CP-odd scenario, respectively. (right) The receiver operating characteristic (ROC) curve, $1 - p_A$ versus $p_h$, of the hypothesis test.}
		\label{q-dist-roc}
	\end{figure}
	
	We perform millions of pseudo-experiments for each of the two scenarios. The pseudo experiment events are taken from the validation set. In FIG.~\ref{q-dist-roc}, we show the probability distributions of the log-likelihood ratio $\ln Q$ from pseudo-experiments (left panel) and receiver operating characteristic (ROC) curves of the hypothesis test (right panel) for different values of integrated luminosity. With the increase of luminosity, we can see that the overlap between the two distributions reduces significantly and the ROC curves will be closer to the corner. When the integrated luminosity reaches 300 fb$^{-1}$, the two distributions will be almost separated. This indicates CP-even and CP-odd components can be well distinguished at the LHC with at most 300 fb$^{-1}$ experimental data.
	
%
	
	
	\section{Acknowledgements}
    We thank the helpful discussions with Dr. Andrew Fowlie.	
	This work was supported by the National Natural Science Foundation of China (NNSFC) under
	grant Nos. 11705093, 11675242, 11821505 and 11851303, 
	by Peng-Huan-Wu Theoretical Physics Innovation Center (11747601),
	by the CAS Center for Excellence in Particle Physics (CCEPP),
	by the CAS Key Research Program of Frontier Sciences
	and by a Key R\&D Program of Ministry of Science and 
	Technology under number 2017YFA0402200-04.
	

	
	%

\end{document}